\documentclass{iau-JDSS}
\usepackage{graphicx}

\title[short title of paper] 
{Molecular lines studies at redshift greater than 1}

\author[short author list]   
{Francoise Combes}

\affiliation{LERMA, Observatoire de Paris,
61 Av. de l'Observatoire, F-75014, Paris \break email: francoise.combes@obspm.fr}

\pubyear{2009}
\volume{Volume 15}  
\pagerange{119--126}
\date{?? and in revised form ??}
\jname{Highlights of Astronomy, Volume 14}
\editors{Ian F Corbett, ed.}
\begin{document}

\maketitle

\begin{abstract}
 Observations of CO molecules in the millimetrer domain
at high redshift (larger than 1), have provided interesting
informations about star formation efficiency, and its
evolution with redshift. Due to the difficulty of
the detections, selection effects are important.
The detection if often due to gravitational amplification.
Objects selected by their (far)infrared flux, are
in general associated to ULIRGS, mergers with starburst
in the nuclear regions. Quasars have been selected
as powerful optical sources, and have been found
to be associated to starbursts, rich in gas.
The gas fraction appears to be much higher at redshift greater than 1.
 Quasars allow to probe the end of the reionisation period,
and the relation between bulge and black hole mass.
However these selection bias could have led us to miss some
gaseous galaxies, with low-efficiency of star
formation, such as the more quiescent objects selected
by their BzK colors at z=1.5 or 2.
\end{abstract}

\firstsection 
\section{Introduction}
 Up to 2009, there is about 50 high redshift sources detected
in the CO lines. This domain of research has grown quickly,
from the first discovery in 1992, of the faint IRAS source
F10214+4724 at=2.3 (Brown \& van den Bout 1992, Solomon et al 1992).
Most of the sources have a redshift between z=2 and 3. Recently, Iono et al (2009)
observed 30 of these sources homogeneously in CO(3-2) with  SMA, and found
a very good correlation between CO and FIR luminosities, even for QSO. This means
that, although the contribution of the AGN to the FIR luminosity increases
with power, the starburst always dominate.

The advantage to detect the CO lines, is to obtain
the efficiency of star formation (ratio of FIR luminosity to
the gas content), which evolves quickly with redshift, and when resolved,
study the kinematics and mass at high z: for example, the source SMM J2399-0136
at z=2.808, has been spatially resolved with the IRAM interferometer, and a hint of
rotation curve has been found at z=2.808 (Genzel et al 2003). Taking into account 
the lens amplification factor of 2.5, the H$_2$ mass derived is
6 10$^{10}$ M$_\odot$, and the dynamical mass is
3.0 10$^{11}$/sin$^2$i  M$_\odot$, uncertain, since the 
inclination of the object is not well determined.

\section{Recent results}

Thanks to strong gravitational amplification, it was possible
to detect CO line emission towards 6 quasars at 
z larger than 4, and determine their surprising properties.

The most distant quasar detected, SDSS J1148+5251 at z=6.4, is a unique
object, at the end of the epoch of reionization. The 
Gunn-Peterson trough is detected in HI absorption in its spectrum
(Fan et al 2003, White et al 2003). In this powerful starburst, of surface
density 1000 M$_\odot$/year /kpc$^2$, an impressive list of molecules
has been detected, and even ionised carbon CII
(158$\mu$m redshifted at 1mm), cf  Walter et al (2009).
Surprisingly the HCN molecule is not detected, while 
HCN is the better tracer of star formation,
well correlated with FIR (Gao \& Solomon 2004).

One of the common features of the z$>$4 quasars, resolved in
the CO(2-1) line with the VLA, is that they are 
gas-rich mergers of galaxies, with complex morphologies, and 
molecular extent of about 5kpc, much larger than local ULIRGs
(Riechers et al 2008a,b).
Once corrected for their strong amplification, their H$_2$ mass
is up to 10$^{11}$ M$_\odot$, a significant fraction of the dynamical
mass. although the latter is ill-defined, given the perturbed shapes.
 Their black hole masses, derived assuming their AGN radiate at Eddington 
luminosity, or from the nuclear emission lines, appear an order of magnitude 
higher than expected from the Magorian relation. But 
uncertainties are large.

 Their star formation surface density is saturating around
1000 M$_\odot$/year /kpc$^2$, as for Eddington limited star formation,
i.e. dust opacity limited gas surface density.
 The average H$_2$ column density reaches
 10$^{24}$ cm$^{-2}$, and the volumic density 10$^{4}$ cm$^{-3}$.

Objects at more moderate SFR are not detected,
unless strongly amplified. Three Lyman-break galaxies (LBG)
 at z$\sim$ 3, such as the Cosmic eye  
(Coppin et al 2007) with a magnification of $\sim$ 30,  also
satisfy the same FIR-CO luminosity relation.
Their star formation rate is of the order of 50  M$_\odot$/year,
with a starburst time-scale of 40Myr, they are the 
high-z analog of local LIRGs.

The most efficient star forming objects appear to be the 
Submillimeter Galaxies (SMG), which have been selected by their
FIR luminosity, redshifted in the mm. They are more efficient than 
ULIRGs, and are very compact starbursts, of radius lower than 1kpc,
or less.  But all detected objects at high z are
not so efficient. Recently, the BzK galaxies, selected for their red colors,
have been detected with much more CO emission than expected, at z=1.5-2
(Daddi et al 2008). These are also ULIRGs, but with a much larger
molecular content, and a time-scale to consume their gas of $\sim$ 2 Gyr.
They are extended 10kpc disks, and their CO excitation is low,
peaking at CO(3-2) like the Milky Way.
Another star forming BzK galaxy however was not detected
by Hatsujade et al (2009), and certainly, they have a wide range of
properties. Due to their low excitation, it might be appropriate
to use a higher CO-to-H$_2$ conversion ratio than for ULIRGs
(Dannerbauer et al 2009).

SMG are much more concentrated, more compact, they are
expected to be major mergers remnants, with low angular momentum,
and precursors of elliptical galaxies (Bouche et al 2007). 
Do they actually trace massive haloes?
In the GOODS-N field, a cluster of radio galaxies and SMG at z=1.99 has
been studied, it is 
the strongest known association of SMG (Chapman et al 2008), with an
overdensity of 10. 
SMG appear also to be associated to filaments traced by Lyman alpha emitters (LAE).
In SSA22, a protocluster region at z=3.1, traces a
filament, where SMG have been detected at 1.1mm with AzTEC on ASTE (Tamura et al 2009).

SMG are also sometimes associated to 
another type of objects, Lyman-alpha blobs (LAB, Geach et al 2007),
huge ionised gas nebulae, excited by a central starburst or AGN.
In local clusters at z=0.4-1, such as CL0024+16, the 
LIRGs detected by Spitzer at 24$\mu$m
can be detected in CO (Geach et al 2009).

The fact that at high redshift, most detected objects are quasars,
allows to study the AGN-starbutst association, and possible
AGN feedback on CO emission. For instance, APM08279+5255 at z=3.9 is a lensed QSO,
(amplification factor ~50),
and one of the brightest object in the sky; it has been observed
with mm and cm telescopes, the CO lines from 
CO(1-0) to CO(11-10) are detected.
Recent 0.3" resolution CO(1-0) mapping with the VLA (Riechers et al 2009),
reveals that the emission is not extended, as previously thought. The amplification 
factor could also be lower.  The CO line emission is
co-spatial with optical/NIR and X-rays, and very compact.
The best model shows that the CO is in a circumnuclear disk of 550 pc radius, 
inclined by 25$^\circ$, with a gas mass of 1.3 10$^{11}$ M$_\odot$.
There is no hint of the influence of the AGN feedback.

\section{Perspectives}

About 50 systems are presently detected in molecular lines at high redshift. 
Given the different properties of the various categories of objects, it becomes obvious that 
results are dominated by selection effects.
 ULIRGs have a very efficient star formation, their molecular gas
has a compact distribution, and is highly excited. But BzK objects have much
more extended gas, with normal star formation efficiency and consumption time-scales.

Present results are also strongly biased by lensing magnification.

One of the main robust results, is that quasars and starbursts are intimately linked,
and the AGN activity does not seem to quench star formation, at these redshifts.


\begin{thebibliography}{}

\bibitem[]{} Bouche N., Cresci, G., Davies, R.  et al.: 2007, ApJ 671, 303 
\bibitem[]{} Brown R.L., van den Bout P.A.: 1992, ApJ 397, L19
\bibitem[]{} Chapman, S. C., Neri, R., Bertoldi, F. et al.: 2008, ApJ 689, 889
\bibitem[]{} Coppin, K. E. K., Swinbank, A. M., Neri, R. et al.: 2007, ApJ 665, 936
\bibitem[]{} Daddi E., Dannerbauer, H., Elbaz, D. et al.: 2008, ApJ 673, L21
\bibitem[]{} Dannerbauer, H., Daddi, E., Riechers, D. A. et al.: 2009, ApJ 698, 178
\bibitem[]{} Fan, X., Strauss, M. A., Schneider, D.P. et al.; 2003, AJ 125, 1649
\bibitem[]{} Gao Y., Solomon P.: 2004, ApJ 606, 271
\bibitem[]{} Geach, J. E., Smail, I., Chapman, S. C. et al.: 2007, ApJ 655, L9
\bibitem[]{} Geach, J. E., Smail, I., Coppin, K. et al: 2009, MNRAS 395, L62
\bibitem[]{} Genzel, R., Baker, A. J., Tacconi, L. et al.: 2003, ApJ 584, 633
\bibitem[]{} Hatsukade, B., Iono, D., Motohara, K. et al.: 2009, PASP 61, 487
\bibitem[]{} Iono, D., Wilson, C. D., Yun, M. S. et al.: 2009, ApJ 695, 1537 
\bibitem[]{} Riechers, D. A., Walter, F., Brewer B. et al.: 2008a, ApJ 686, 851 
\bibitem[]{} Riechers, D. A., Walter, F., Carilli, C. et al.: 2008b, ApJ 686, L9
\bibitem[]{} Riechers, D. A., Walter, F., Carilli, C. et al.: 2009, ApJ 690,  463 
\bibitem[]{} Solomon P., Radford, S. J. E., Downes, D.: 1992, Nature 356, 318
\bibitem[]{} Tamura, Y., Kohno, K., Nakanishi, K. et al.: 2009, Nature 459, 61
\bibitem[]{} Walter F., Riechers, D., Cox, P. et al.: 2009, Nature 457, 699
\bibitem[]{} White, R. L., Becker, R. H., Fan, X., Strauss, M. A.: 2003, AJ 126, 1


\end{thebibliography}
\end{document}